\documentclass[aps,prl,floatfix,superscriptaddress,titlepage,twocolumn,amsmath,amssymb,nofootinbib]{revtex4}

\usepackage{graphicx}
\usepackage{url}
\usepackage{amsmath}
\usepackage{amssymb}
\usepackage{bm}
\usepackage{dsfont}
\usepackage{calc}
\usepackage{array}
\usepackage{courier}
\usepackage{scalefnt}
\usepackage{braket}
\usepackage{times}
\usepackage{units}
\usepackage{yfonts}
\usepackage{calc}
\usepackage{color}
\usepackage{braket}
\usepackage[normalem]{ulem} 

\bibpunct{}{}{,}{s}{}{\textsuperscript{,}}

\renewcommand{\section}[1]{~\\{\fontsize{10.5pt}{14pt}\textbf{#1}}\\}

\renewcommand{\subsection}[1]{~\\\normalsize\noindent{\fontsize{9.5pt}{14pt}\textbf{#1}}\\}

\newcommand{\methodsection}[1]{\small\noindent{\textbf{#1.}}~} 

\definecolor{darkblue}{rgb}{0.2,0.2,0.8}
\usepackage[pdftex,
        colorlinks=true,
        linkcolor=darkblue,
        citecolor=darkblue,
        filecolor=black,
        urlcolor=darkblue,
        bookmarks=false,
        bookmarksopen=false,
        bookmarksopenlevel=3,
        plainpages=false,
        pdfpagelabels=true,
        verbose=true]{hyperref}
\usepackage[all]{hypcap}
\newcommand{\Methods}{\hyperlink{MethodSection}{Methods}}

\newcommand{\expect}[1]{\langle {#1} \rangle}
\newcommand{\eff}{\mathrm{eff}}
\newcommand{\up}{\uparrow}
\newcommand{\down}{\downarrow}

\newcommand{\e}{\mathrm{e}}
\newcommand{\I}{\mathrm{i}}
\renewcommand{\:}{\,$$=$$\,}

\newcommand{\ER}{E_\mathrm{R}}

\newcommand{\Fig}[1]{Fig.~\ref{#1}}

\newcommand*\phantomas[3][c]{\ifmmode \makebox[\widthof{$#2$}][#1]{$#3$}\else \makebox[\widthof{#2}][#1]{#3}\fi}
\newcommand{\phiA}{\phi_\mathrm{A}}
\newcommand{\phiB}{\phi_\mathrm{B}}

\newcommand{\citeMI}{\cite{Fisher1989,Jaksch1998,Greiner2002}}
\newcommand{\citePSF}{\cite{Kuklov2004,Mathey2009,Hu2009,Menotti2010,Sowinski2012,Wang2013,Zhang2013,Zhu2014}}
\newcommand{\citePSS}{\cite{Mathey2009,Hu2009,Wang2013,Zhang2013}}
\newcommand{\citeCDW}{\cite{Kuhner1998,Kuhner2000,Sengupta2005,Mazzarella2006,Wessel2007,Gan2007a,Gan2007b,Mathey2009,Ng2010,Iskin2011,Kurdestany2012,Maik2013}} 
\newcommand{\citeSS}{\cite{Andreev1969,Chester1970,Leggett1970,Batrouni2000,Sengupta2005,Mazzarella2006,Wessel2007,Gan2007a,Gan2007b,Mathey2009,Ng2010,Kurdestany2012,Maik2013}}
\newcommand{\citeDMIExclUs}{\cite{Buonsante2004b,Buonsante2005a,Buonsante2005b,Danshita2007,Muth2008}}
\newcommand{\citeDMI}{\cite{Buonsante2004b,Buonsante2005a,Buonsante2005b,Danshita2007,Muth2008,Jurgensen2014}}
\newcommand{\citeCFSF}{\cite{Kuklov2003,Altmann2003,Kuklov2004,Hu2009,Hubener2009,Ozaki2012}}

\begin{document}

\title{Twisted complex superfluids in optical lattices}

\author{Ole J\"urgensen}
\author{Klaus Sengstock}
\author{Dirk-S\"oren L\"uhmann}

\affiliation{Institut f\"ur Laser-Physik, Universit\"at Hamburg, Luruper Chaussee 149, 22761 Hamburg, Germany}

\maketitle

\onecolumngrid{\vspace{-1em}\bf\noindent
We show that correlated pair tunneling drives a phase transition to a twisted superfluid with a complex order parameter. 
This unconventional superfluid phase  spontaneously breaks the time-reversal symmetry and is characterized by a twisting of the complex phase angle between adjacent lattice sites. We discuss the entire phase diagram of the extended Bose--Hubbard model for a honeycomb optical lattice showing a multitude of quantum phases including twisted superfluids, pair superfluids, supersolids and twisted supersolids. Furthermore, we show that the nearest-neighbor interactions lead to a spontaneous breaking of the inversion symmetry of the lattice and give rise to dimerized density-wave insulators, where particles are delocalized on dimers. For two components, we find twisted superfluid phases with strong correlations between the species already for surprisingly small pair-tunneling amplitudes. Interestingly, this ground state shows an infinite degeneracy ranging continuously from a supersolid to a twisted superfluid.
}\\\twocolumngrid


The time-reversal symmetry of the Schr\"odinger equation allows us to describe superfluid ground states with real wave functions in accordance with Feynman's no-node theorem \cite{Feynman1972}. This principle holds unless the Hamiltonian breaks this symmetry either explicitly or spontaneously, which is typically the case for higher orbitals or spin-orbit coupling \cite{Wu2009}. Recently, experiments with binary mixtures of ultracold bosonic atoms in a honeycomb optical lattice have observed a multi-orbital superfluid \cite{Soltan-Panahi2012}. Most strikingly, time-of-flight measurements have revealed a complex superfluid order parameter. The complex phase angle of the superfluid order parameter twists between neighboring lattice sites without accumulating  a total flux. Furthermore, a phase transition between a normal and this so-called twisted superfluid was observed. The origin of this unconventional superfluid phase lacks a conclusive theoretical understanding. In fact, previous theoretical studies \cite{Choudhury2013,Cao2014} have found no indication for a transition to the twisted superfluid phase. Here, we show that correlated pair tunneling can drive the quantum phase transition to the twisted superfluid ground state. This quantum phase is remarkable in several ways.  It spontaneously breaks the time-reversal symmetry, combines off-diagonal long-range order with diagonal short-range order and is crucially based on beyond-mean-field correlations.  Interestingly, the twisted multi-orbital superfluid has a strong connection to mechanisms discussed for superconducting materials. Two-orbital order parameters have also been experimentally observed \cite{Sun1994, Kouznetsov1997, Kaminski2002, Lu2008} in high-temperature superconductors and play a central role for the understanding of this phenomenon. A spontaneous time-reversal symmetry breaking associated with complex order parameters has been observed in the pseudo-gap phase of the high-temperature superconductor Bi-2212 with ARPES \cite{Kaminski2002}, as well as in $\mathrm{Sr}_2\mathrm{RuO}_4$ using a muon spin relaxation measurement \cite{Luke1998}  and $\mathrm{UPt}_3$ via the polar Kerr effect \cite{Sauls1994, Strand2010, Schemm2014}. In this respect, the twisted superfluidity interlinks many-body quantum gas systems with superconducting materials.
                                    
The Hubbard model is a primary description for strongly correlated electrons in lattices. It accounts for the single-particle tunneling between neighboring sites with amplitude $J$ and the interaction $U$ of a pair of particles on the same lattice site. Its counterpart for bosonic particles is the Bose--Hubbard model. For weak interactions the ground state is a superfluid phase (SF), whereas for strong interactions a localization of the particles is favored forming the Mott insulator phase (MI) \citeMI. Theoretically, extended Bose--Hubbard models with off-site interactions \cite{Dutta2015} have been studied predicting pair superfluid phases \citePSF, charge-density-wave insulators \citeCDW~and supersolid phases \citeSS. The latter two phases exhibit a density modulation with respect to neighboring lattice sites and break the inversion symmetry of the lattice Hamiltonian spontaneously in the thermodynamic limit \cite{Kuhner1998,Kuhner2000}. 

\begin{figure*}
\centering
\includegraphics[width=1\linewidth]{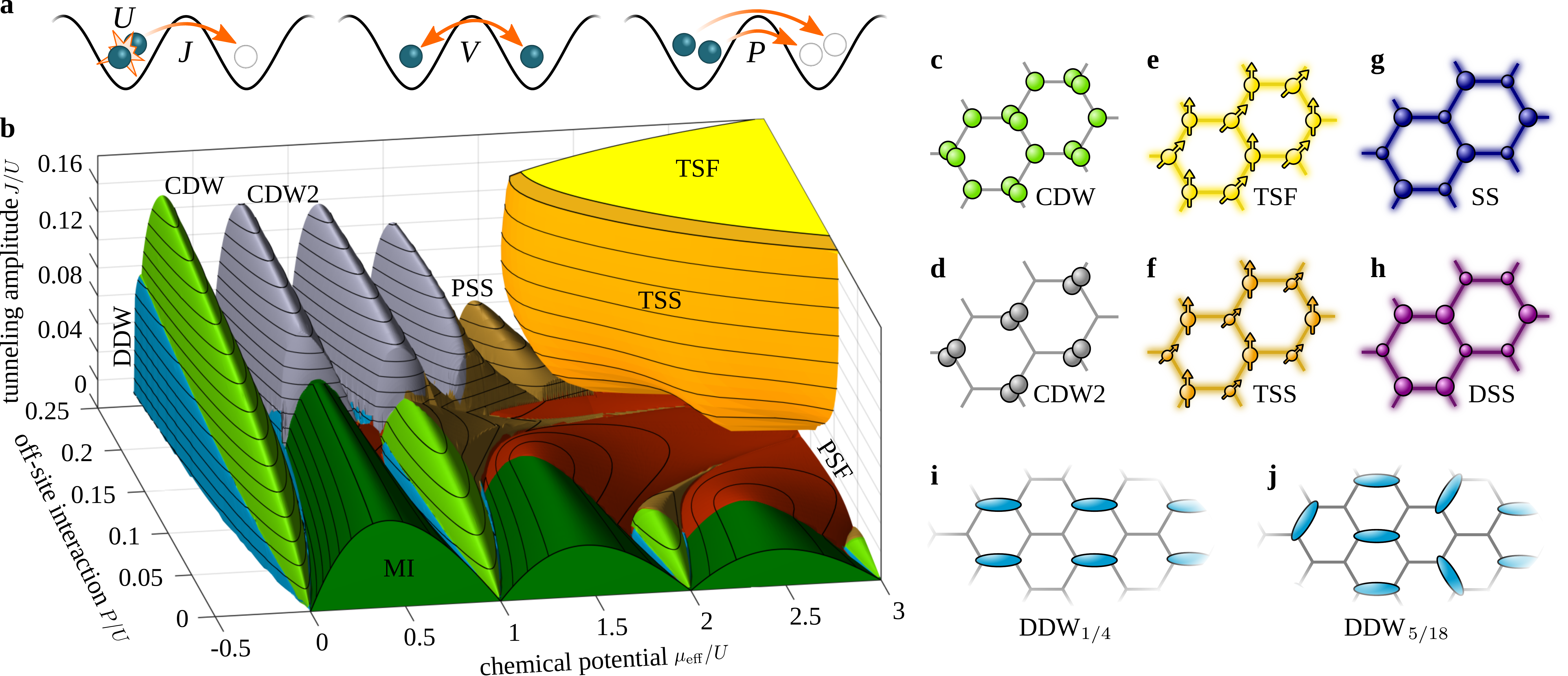}
\caption{\textbf{Phase diagram of the single-component extended Bose--Hubbard model.} \textbf{a} Schematics for the processes in the extended Bose--Hubbard model. \textbf{b} Critical value of $J/U$ for the transition to the superfluid (or supersolid) phase as a function of the effective chemical potential $\mu_\eff/U$ and the off-site interaction amplitudes $P/U=V/U$ of pair tunneling and nearest-neighbor interaction. The three-dimensional phase diagram is obtained using a two-site cluster. The colors indicate the distinct quantum phases. Insulator phases with integer site occupancy are the Mott insulator (MI, dark green), \textbf{c} the charge-density-wave insulator (CDW, light green), and \textbf{d} the maximally-imbalanced CDW insulator (CDW2, gray). The phases \textbf{e}-\textbf{h} have a non-vanishing superfluid order parameter (indicated by a blur). \textbf{e} The twisted superfluid (TSF, yellow) is characterized by a complex order parameter $\expect{\hat a_i}$, where the arrows indicate the complex phase angle $\arg\expect{\hat a_i}$. \textbf{f} The twisted supersolid (TSS, orange) phase has an additional density wave. \textbf{g} Supersolids (SS) combine a real order parameter with a density wave (not shown in \textbf{b}). \textbf{h} Dimerized supersolids (DSS) are characterized by a long-range density wave with no density gradient within unit cells.  \textbf{i} At quarter integer filling, dimerized density-wave insulators (DDW, blue) appear where particles are delocalized on two sites. \textbf{j} In addition, dimerized insulators with other fractional fillings such as $5/18$ are possible. For large values of $P$ and $\mu$, pair superfluids (PSF, red) and pair supersolids (PSS, brown) can be found. 
\label{fig_pd}}   
\end{figure*}

We study an extended Bose--Hubbard model with nearest-neighbor interaction and pair tunneling on a honeycomb lattice using cluster-Gutzwiller theory. This method takes into account short-range correlations exactly and simultaneously describes the thermodynamic limit of infinite two-dimensional lattices. The extended Hubbard processes give rise to several novel correlated quantum phases. First, we discuss the driving mechanism behind the twisted superfluid phase and the spontaneous breaking of the time-reversal symmetry. Second, we show how nearest-neighbor interaction supports dimerized density waves, where the particles are delocalized on dimers and the translational symmetry is broken. This insulator is surrounded by a dimerized supersolid exhibiting the same density modulation. Third, we extend our study to a quantum gas mixture of two spin components. Remarkably,  the two-species twisted superfluidity persists even at small pair-tunneling amplitudes allowing its experimental realization with ultracold atoms \cite{Soltan-Panahi2012}. The twisted superfluid is characterized by strong correlations between the two components and exhibits an infinite ground-state degeneracy.

\section{Results}
We study the extended Bose--Hubbard model with two additional processes arising from the interaction of particles on adjacent sites known as nearest-neighboring interaction $V$ and (correlated) pair tunneling $P$ (see \Fig{fig_pd}a). As in the experimental realization \cite{Soltan-Panahi2012}, we consider a honeycomb lattice geometry. First, we restrict our calculations to bosonic particles with a frozen spin degree of freedom. For a single component, the Hamiltonian of the extended Bose--Hubbard model reads   
\begin{equation}\begin{split}\label{eq_Hfull}
 \hat H = &\frac{U}{2} \sum_i \hat n_i (\hat n_i - 1) -J \sum_{\langle i,j \rangle} \hat a^\dagger_i \hat a_j - \mu \sum_i \hat n_i \\
+&V \sum_{\langle i,j \rangle} \hat n_i \hat n_j + \frac{P}{2} \sum_{\langle i,j \rangle} \hat a^\dagger_i \hat a^\dagger_i \hat a_j \hat a_j,
\end{split}\end{equation}
where the operators $\hat a_i$ ($\hat a^\dagger_i$) annihilate (create) a particle on site $i$ and $\hat n_i = \hat a^\dagger_i\hat a_i$. The brackets $\langle i, j \rangle$ indicate pairs of nearest neighbors $i$ and $j$. The first line of the Hamiltonian is the standard Bose--Hubbard model with on-site interaction energy $U$, tunneling amplitude $J$ and chemical potential $\mu$. The second line describes off-site interaction processes $V$ and $P$ giving rise to the exceedingly rich quantum phase diagram shown in \Fig{fig_pd}b. For neutral ultracold atoms, it is valid to assume a contact interaction potential. As a consequence, the amplitudes of the nearest-neighbor interactions $V$ and the pair tunneling $P$ are equal. 

The phase-diagram in \Fig{fig_pd}b is obtained by means of a cluster-mean field theory. Within each cluster quantum-mechanical correlations are treated exactly whereas the boundary is coupled self-consistently to the mean field. Unlike in the standard Hubbard model, the respective factorization of the Hamiltonian \eqref{eq_Hfull} leads to three distinct mean fields, namely $\expect{\hat a_i }$, $\expect{\hat n_i}$ and $\expect{\hat a_i \hat a_i}$ (see \Methods ). The increased complexity limits the numerical treatment to clusters of up to four sites. The phase diagram \Fig{fig_pd}b covers the whole three-dimensional parameter space of the Hamiltonian \eqref{eq_Hfull} for $P\:V$. For small ratios $J/U$ insulator phases dominate the phase diagram except for the pair superfluid (PSF) {\citePSF}~and pair supersolid (PSS) {\citePSS}~where the order parameter $\expect{\hat a_i }$ vanishes but the pair order parameter $\expect{\hat a_i \hat a_i}$ takes a finite value.

\subsection{Twisted complex superfluidity}
The most striking feature is the appearance of the twisted superfluid phase (TSF).  As illustrated in \Fig{fig_pd}e, this unconventional superfluid is characterized by a twist of the complex phase angle of the superfluid order parameter between neighboring sites, whereas the density $\expect{\hat n_i}$ is homogeneous. Expressing the local order parameter as $\expect{\hat a_i}\:\phi_i \e^{\I \theta_i}$, the difference of the complex phase angle of adjacent sites is {$\theta\: \theta_i$$-$$\theta_j$}. The ground state breaks the time-reversal symmetry and is two-fold degenerate with a twisting of either $\theta$ or $-\theta$. The two degenerate many-particle states $\{\psi,\mathrm{T}\psi\}$ are connected by the time-reversal symmetry operation $\mathrm{T}\psi\:\psi^*$. In Ref.~\citenum{Soltan-Panahi2012}, the complex superfluid ground state has been observed experimentally. It has been linked to the emergence of a two-mode superfluid order parameter on the basis of a simple variational mean-field model. Previous theoretical studies have found only real-valued superfluid ground states \cite{Choudhury2013,Cao2014}. These approaches either discard correlations \cite{Choudhury2013} or are restricted to finite-size systems and the standard Bose--Hubbard model \cite{Cao2014}. In a recent work on a double well using a mean-field model a complex ground state has been found for large pair-tunneling amplitudes \cite{Zhu2014}. Our theoretical framework allows a correlated treatment of a lattice in the thermodynamic limit incorporating beyond-Hubbard processes and offers a conclusive understanding of the twisted superfluid quantum phase.

At first glance, the complex order-parameter and the time-reversal symmetry breaking seem to stand in contradiction to Feynman's no-node theorem \cite{Feynman1972,Wu2009}, which we resolve in the following. For real Hamiltonians, the no-node theorem implies a positive real and therefore non-degenerate ground state. However, the energy difference between the lowest-energetic states can become arbitrarily small with increasing system size. It has been shown that the energy gap between the lowest two states vanishes in the thermodynamic limit for the charge density wave \cite{Kuhner2000} as well as for a highly occupied double well \cite{Zhu2014}. In the first case, the superposition of two opposite density waves represents the ground state in finite systems but a small perturbation breaks the inversion symmetry in favor of one density wave. In the second case, the time-reversal symmetry becomes unstable against perturbations for a large number of particles. The mean-field approach reproduces the thermodynamic limit and tends to break both symmetries via the incoherent mean-field coupling at the boundary of the cluster. For real macroscopic systems, such incoherences are introduced by several mechanisms such as system-bath coupling with the surrounding, thermodynamic excitations, fragmentation, infinite relaxation times and dissipation in open systems.    
     
The driving mechanism of the twisted superfluid phase is the correlated pair tunneling $P$ competing with the single-particle tunneling $J$. The competition becomes apparent when introducing simple variational mean-field expressions. Replacing all operators $\hat a_i$ with complex numbers $\phi_i \e^{\I \theta_i}$ (see \Methods), we can approximate both contributions as 
\begin{equation}\begin{split}\label{eq_simple_mf}
 -J \expect{\hat a^\dagger_i \hat a_j } + \mathrm{c.c.}\approx&-2J \phi_i \phi_j \cos(\theta) , \\
P/2\ \expect{\hat a^\dagger_i \hat a^\dagger_i \hat a_j \hat a_j} + \mathrm{c.c.} \approx & \ P \phi_i^2 \phi_j^2 \cos(2\theta)
\end{split}\end{equation}
with $\theta\:\theta_i$$-$$\theta_j$. For small amplitudes of the pair tunneling $P$, the first term dominates and minimizing the energy yields $\theta\:0$. The pair tunneling energy on the other hand is minimized for $\theta\:\pm \pi/2$. Due to the different functional dependence on $\theta$, the system undergoes a phase transition from $\theta\:0$ to a finite value $|\theta|$$>$$0$ at a critical ratio $P/J$. In a more qualitative picture, we have to include correlations that are strong near the insulating phases, i.e.~$\expect{\hat n_i}$ and $\expect{\hat a_i \hat a_i}$ may differ strongly from the variational mean-field expressions $\phi_i^2$ and $\phi_i^2 e^{2\I \theta_i}$, respectively. This is accounted for in the applied cluster Gutzwiller approach treating the cluster sites in a full many-particle description. Additionally, the off-site interaction $V$ creates a competition between the phase twist and density modulations. The interplay of insulators, density waves and complex twisting of the order parameter is vital to the phase diagram in \Fig{fig_pd}b. As a result, a multitude of quantum phase transitions appear which we discuss in the following.   

\begin{figure}
\centering
 \includegraphics[width=\linewidth]{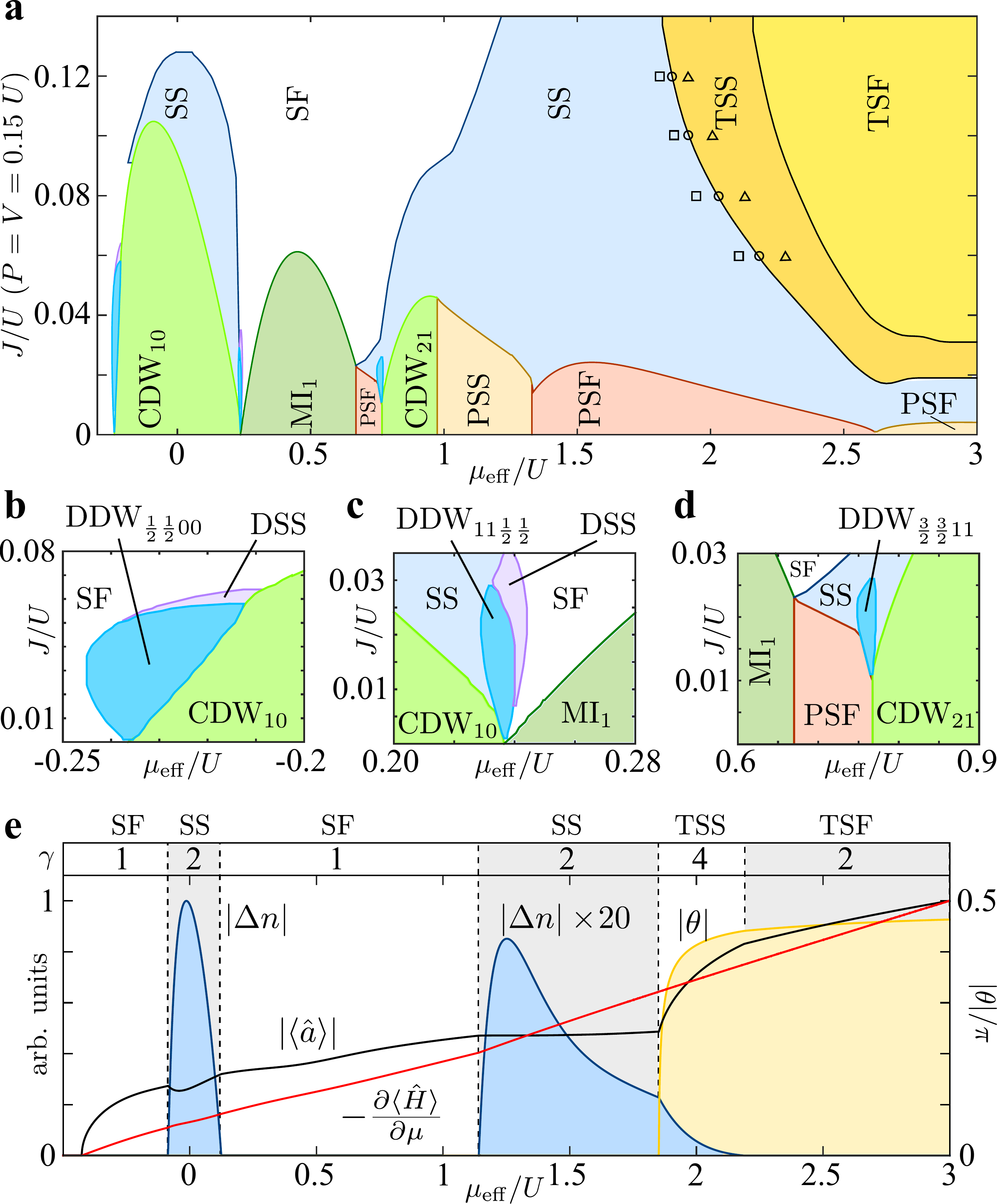}
\caption{\textbf{Phase boundaries and order parameters}. \textbf{a} Cut through the phase diagram \Fig{fig_pd}b showing the plane with $P\:V\:0.15\, U$ (see \Fig{fig_pd}b for abbreviations). The markers indicate the critical point for the SS-TSS phase transition obtained for cluster sizes 1 (triangle), 2 (circle) and 4 (square). \textbf{b-d} Details of the phase-diagram showing dimerized density-wave phases (DDW) with fillings $1/4$, $3/4$ and $5/4$ as well as supersolids with long periodicity (DSS). \textbf{e} Averaged superfluid order parameter $|\expect{\hat a}|$ (black), its complex phase angle $\theta$ (yellow), density wave $|\Delta n|$ (blue), and the energy derivative (red) shown across various phase transitions for $J/U=0.12$, where $\gamma$ is ground-state degeneracy. 
\label{fig_pd_cut}}
\end{figure}

Figure~\ref{fig_pd_cut}a shows a cut through the phase diagram for $P\:V\:0.15U$, which depicts also the supersolid phase (SS) omitted in \Fig{fig_pd}b. The supersolid has a non-vanishing superfluid order parameter combined with a density wave representing diagonal order \citeSS (\Fig{fig_pd}g). The respective expectation values for $\expect{\hat a_i}$ and $\Delta n\:\expect{\hat n_i- \hat n_j}$ are shown for $J\:0.12$ in \Fig{fig_pd_cut}e. In the thermodynamic limit, the supersolid phase breaks the inversion symmetry $\mathrm{P}$ and has a two-fold degenerate ground state $\{\psi, \mathrm{P}\psi\}$, where the operation $\mathrm{P}$ interchanges the two hexagonal sublattices A and B. 

The transition between the supersolid and the twisted superfluid occurs via a twisted supersolid (TSS) which combines the properties of both phases. In the twisted supersolid both the density wave and the complex twisting $\theta\!\neq\!0$ of the order parameter occur (\Fig{fig_pd}f). Both inversion and time-reversal symmetry are broken and the ground state is four-fold degenerate $\{\psi, \mathrm{P}\psi, \mathrm{T}\psi, \mathrm{PT}\psi\}$. The superfluid order parameter, density wave and phase twist $\theta$ have a kink at the SS-TSS as well as at the TSS-TSF transition (\Fig{fig_pd_cut}e). In the TSS phase the phase twist $\theta$ increases rapidly towards $\pm\pi/2$ with increasing $\mu_\eff$, whereas the density modulation drops to zero when entering the TSF phase. This means that the complex phase of the order parameter suppresses the density wave reflecting the interplay between nearest-neighbor interaction $V$ and pair tunneling $P$.  The energy derivative $-\partial \hat H/\partial \mu $ has no discontinuities but kinks at transitions between SF and SS phases marking all phase transitions as second (or higher) order. 
For the transition from the TSS to the SS, we can evaluate the energy gap $\Delta$ between the ground state and the real-valued SS state. At the phase boundary, we find the critical behavior $\Delta \propto |\mu-\mu_c|^{z\nu}$,  $|J-J_c|^{z\nu}$ and  $|P-P_c|^{z\nu}$ with a critical exponent $z\nu=1.96\pm0.10$. 
We observe that the TSS phase boundary expands when increasing the cluster size of the calculation (\Fig{fig_pd_cut}a) indicating that finite-size scaling \cite{Luhmann2013} would lead to an even larger region where the superfluid order parameter is complex valued. Note that \Fig{fig_pd}b is a zero-temperature phase diagram. We expect the energetically lowest excitation within the twisted superfluid phase to be a disturbance of the phase angle, which would be associated with an energy on the order of the pair tunneling $P$.

\subsection{Dimerized density-wave insulators}
For small $J/U$ several insulator phases can be found with a vanishing order parameter $\langle \hat a \rangle$ (Figs.~\ref{fig_pd} and \ref{fig_pd_cut}). For $V\:P\:0$, we find the usual Mott insulators (MI) \citeMI~with integer filling $\expect{\hat n_i}\:1,2,3$ on A and B sites. For increasing off-site interaction, charge-density-wave insulators (CDW) \citeCDW~with half-integer filling per lattice site appear in between the Mott lobes. These incompressible phases are characterized by an alternating filling $n$ and $n+1$ on A and B sites (\Fig{fig_pd}c). Like in the supersolid phases, the inversion symmetry of the lattice is spontaneously broken in the thermodynamic limit. Responsible for this is the nearest-neighbor interaction $V$ which is reduced for an alternating filling. The charge-density wave phases are also referred to as Mott solids, alternating Mott insulators, and for square lattices as checkerboard insulators. 

Remarkably, in between charge-density waves and Mott insulators, we also find fractional insulator phases with quarter-integer filling indicated in blue in Figs.~\ref{fig_pd} and \ref{fig_pd_cut}. These dimerized density waves (DDW) combine a density wave and a vanishing superfluid order parameter with a spatial delocalization of particles on dimers of the lattice. Similar phases  have been discussed only for lattices with non-symmetric tunneling and are referred to as "loophole" or fractional insulators \citeDMIExclUs~or in the context of honeycomb lattices as dimerized Mott insulators \cite{Jurgensen2014}. For translational-symmetric lattices this kind of dimerization without a density wave is entirely unexpected. We find that for non-zero nearest-neighbor interaction $V$ dimerized phases with  quarter-integer filling can emerge from the spontaneous breaking of the translational symmetry. For a filling of $1/4$, a long-range density wave emerges with a delocalized particle on two sites alternating with two empty sites (illustrated in \Fig{fig_pd}i). On these four sites, the respective many-particle state can be written approximately as $(\,\ket{0,1,0,0} + \ket{1,0,0,0}\,) / \sqrt{2}$. Due to six possible configurations on the honeycomb lattice, this state is six-fold degenerate. Like the CDW, the dimerized density-wave minimizes the $V$ energy but also allows the particles gaining energy from the single-particle tunneling $J$. The first-order energy $E_{\mathrm{DDW}_{1/4}}\: -J-\mu$ per unit cell can fall below the energy of the CDW $E_{\mathrm{CDW}_{10}} \: - 2\mu$.

 The close-ups of the DDW phases for filling $1/4$, $3/4$ and $5/4$ in \Fig{fig_pd_cut}b-d reveal the typical droplet shape that is also present for non-symmetric lattices \citeDMI~as well as a direct insulator-insulator transition in \Fig{fig_pd_cut}b. We also find this kind of dimerization in the supersolid state. Close to the dimerized density-wave with filling $1/4$ and $3/4$, a phase emerges with a non-vanishing superfluid order parameter and equal densities on neighboring dimer sites but a long-range density wave (\Fig{fig_pd_cut}b,c). It is denoted as dimerized supersolid (DSS) and illustrated in \Fig{fig_pd}h.

 In the dimerized density-wave, neighboring sites are strongly correlated and thus single-site mean-field approaches are unable to find this quantum phase.  Due to the cluster size for this calculation, filling at multiples of $n=1/4$ are expected, but depending on the lattice geometry, different fillings might be possible and energetically favorable depending on the chemical potential. For example, the structure shown in \Fig{fig_pd}j has the same energy per particle as the DDW$_{1/4}$ state but with a larger filling $5/18$ indicating that fragmentation into larger unit cells might be possible. 
 
Before turning to two-component systems, we briefly discuss other phases appearing in Figs.~\ref{fig_pd} and \ref{fig_pd_cut} that are already discussed in the literature. At a critical value $V=U/6$ it becomes energetically favorable to gather all particles on one sublattice, while for smaller values of $V$ the population imbalance between the sites is minimized \cite{Iskin2011}. As a result, both Mott and CDW insulators undergo the transition to a maximally-imbalanced charge-density wave (CDW2). For large values of $V$, the dimerized density waves (with filling $\geq 3/4$) have a smooth crossover to a long-range density-wave with maximal imbalance between neighboring sites, i.e. $\ket{n,0,n-1,0}$. For sufficiently large values of $\mu_\eff$ the insulators develop a finite pair superfluidity $\langle \hat a^2 \rangle$ driven by the pair tunneling $P$. Figure \ref{fig_pd}b shows that pair superfluids (PSF) \citePSF~extend from the Mott insulator phases, whereas CDW insulators are attached to pair supersolids (PSS) \citePSS~characterized by both a density wave and pair superfluidity.  

\begin{figure}
\centering
 \includegraphics[width=1\linewidth]{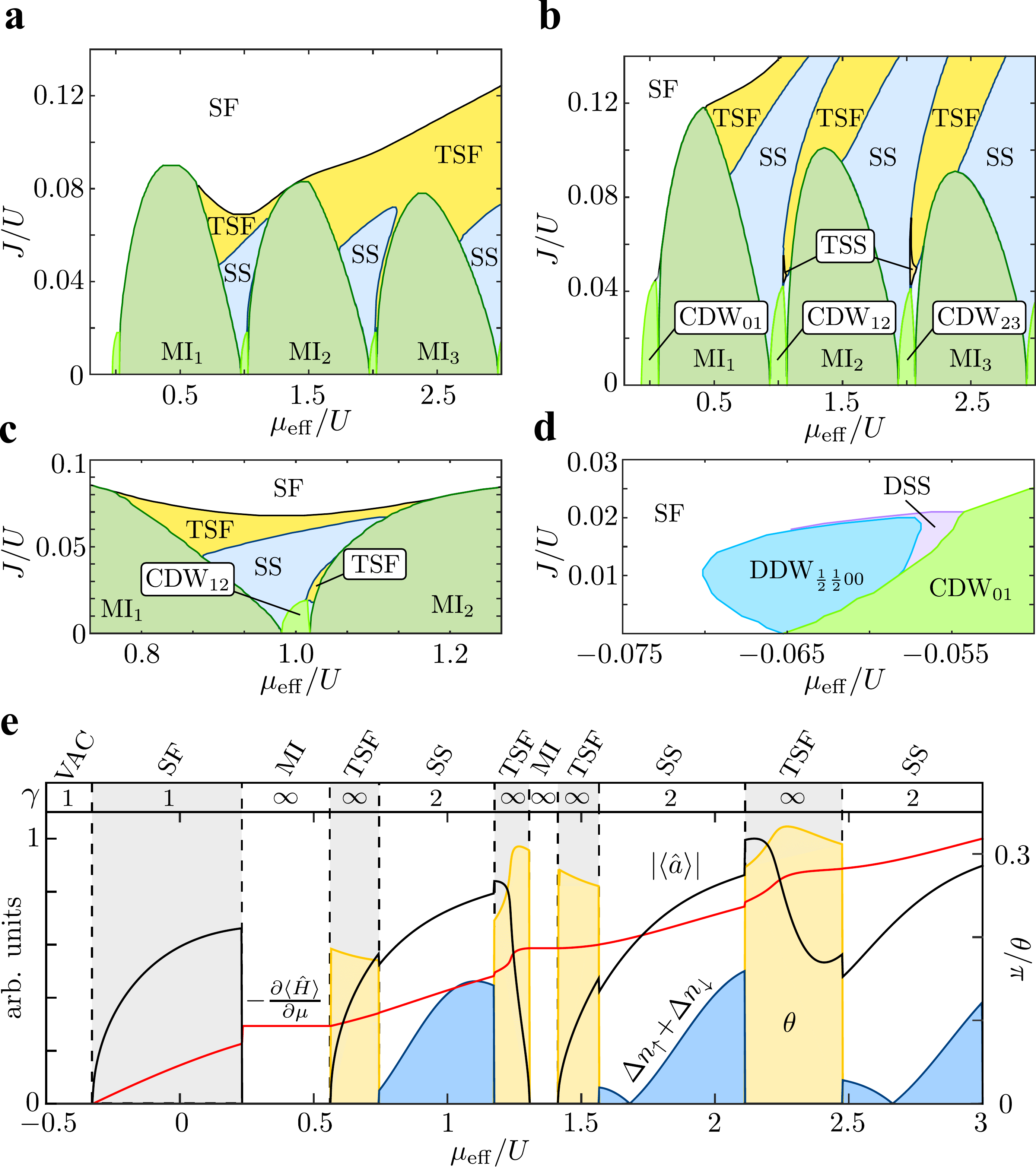}
\caption{\textbf{Phase diagrams of the extended two-component Bose--Hubbard model.} The results are obtained for \textbf{a} $P\:0.02\, U$ and \textbf{b} $P\:0.05\, U$ with $P\:V\:C$ using a single-site cluster. The close-ups below for \textbf{c} $P\:0.02\, U$ and \textbf{d} $P\:0.05\, U$ are computed with a two-site cluster. This allows identifying dimerized density waves (DDW) and dimerized supersolids (DSS). \textbf{e} Order parameters and degeneracy $\gamma$ for $J\:0.1\, U$ and  $P\:0.05\, U$. In the continuously degenerate phases the expectation values correspond to the solution with $\theta\:\theta_\mathrm{max}$.  \label{fig_pd2species}}
\end{figure}

\subsection{Twisted superfluidity for two components}
In the following, we discuss the quantum phases described above for a system with two distinguishable bosonic species. This introduces a new degree of complexity and has a dramatic impact on the phase diagram. The two bosonic components, denoted as $\up$ and $\down$, can e.g.~be realized as atoms in two different hyperfine states. For simplicity, we assume symmetric intra- and interspecies interactions as well as an inversion-symmetric honeycomb lattice. Note that in the experiment \cite{Soltan-Panahi2012} the honeycomb lattice is strongly spin-dependent with a site-offset between the sublattices. Furthermore, we assume a balanced population of both species, i.e $\sum_i \expect{\hat n_{\up i}} = \sum_i \expect{\hat n_{\down i}}$. The two-component Hamiltonian is given by 
\begin{equation} \label{eq_H_2species}
\hat H=\hat H_\up + \hat H_\down + U\sum_i \hat n_{\up i} \hat n_{\down i} +\hat H_{\up\down}^\mathrm{nn}, 
\end{equation}
where $\hat H_\up$ and $\hat H_\down$ are the intraspecies Hamiltonians \eqref{eq_Hfull}. The nearest-neighbor interspecies interaction reads
\begin{equation} \label{eq_H_2speciesUD}
\hat H_{\up\down}^\mathrm{nn}\! =\! \sum_{\langle i,j \rangle}  V\, \hat n_{\up i} \hat n_{\down j} 
+  P\,  \hat a^\dagger_{\up i} \hat a^\dagger_{\down i} \hat a_{\down j} \hat a_{\up j} 
+ C\, \hat a^\dagger_{\up i} \hat a^\dagger_{\down j} \hat a_{\down i} \hat a_{\up j}
\end{equation}
and accounts for the nearest-neighbor interaction $V$, the tunneling of a $\up,\down$-pair with amplitude $P$ and the cross tunneling (or counter hopping) $C$, which interchanges a $\up$- and a $\down$-particle on neighboring sites. For spin-independent contact interaction, the amplitudes  $V\:P\:C$ are equal to the corresponding intraspecies interactions. The complexity of the two-component Hamiltonian is reflected by eight independent mean-field parameters that are needed for the description (see \Methods). In addition, the number of the many-particle states in the cluster increases drastically by incorporating the tensor basis of the two components limiting the cluster size to maximally two sites.  
 
The phase diagram for the binary bosonic mixture is shown in \Fig{fig_pd2species} and differs strongly from the single-component case. Surprisingly, the twisted superfluid phase (TSF) appears already for small values of the chemical potential and, most notably, for small amplitudes of the pair tunneling. For $P\:0.05U$, the TSF phases form stripes in the phase diagram emerging from the charge density waves (CDW) (\Fig{fig_pd2species}b) and are separated by supersolid phases (SS). For a smaller amplitude $P\:0.02U$ the stripes overlap as shown in \Fig{fig_pd2species}a. We observe that the slope of the SF-TSF phase boundary is roughly proportional to $P$. To first order, this correspond to a constant ratio $\expect{\hat P}/\expect{\hat J}\approx P \expect{\hat n}^2/ J \expect{\hat n}$ leading to $J\propto\expect{\hat n} P$.     
 
\begin{figure}
\centering
 \includegraphics[width=1\linewidth]{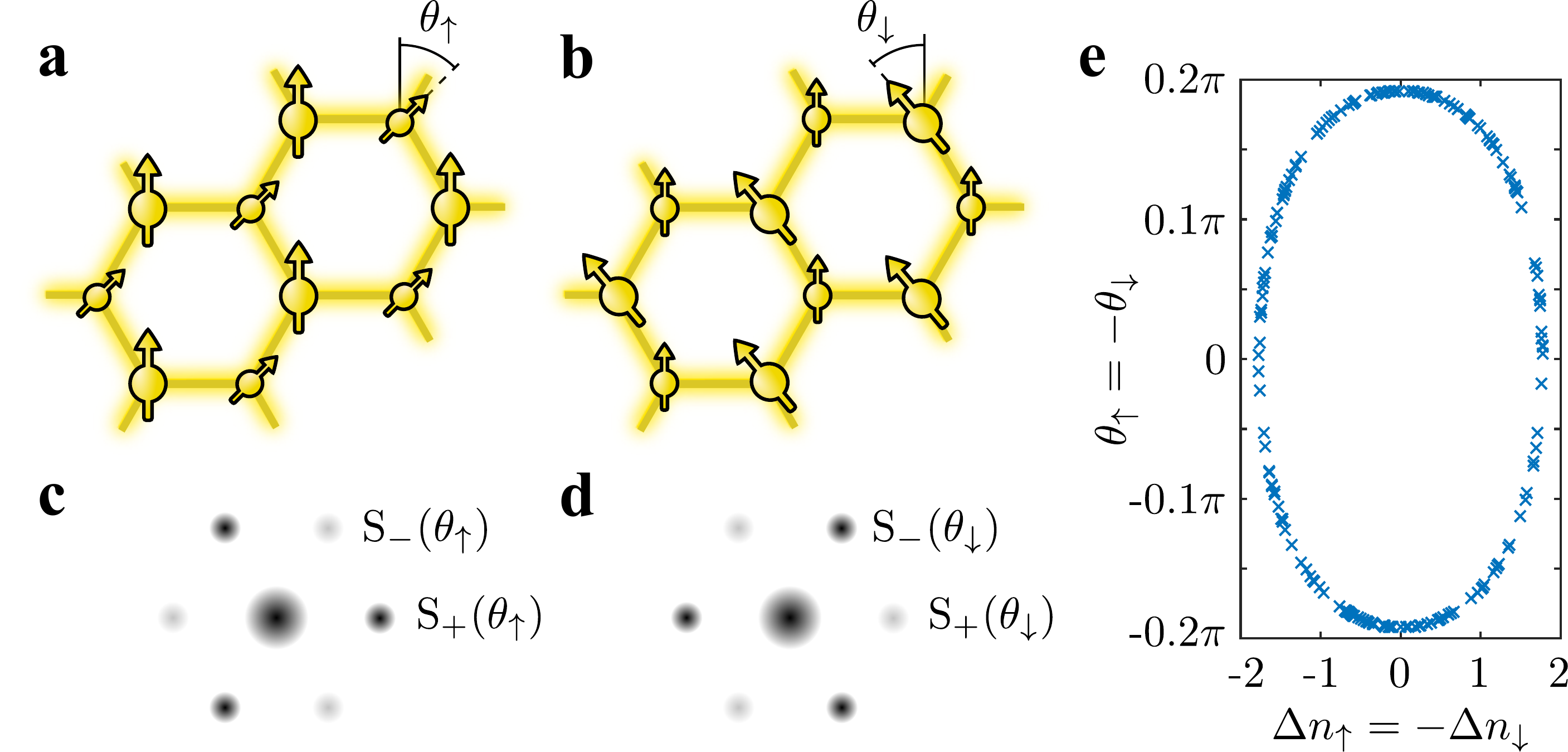}
\caption{\textbf{Experimental signature of  twisted superfluidity.} For two components, denoted as $\up$ and $\down$, the twisted superfluid phase is characterized by an anti-correlation of phase twist and density wave. \textbf{a} The arrows illustrate a complex phase angle $\theta_\up\:\theta$ for the $\up$-component. \textbf{b} The $\down$-component has the opposite twist $\theta_\down\:-\theta$. The density waves oppose each other forming a spin-density wave with a homogeneous total density. \textbf{c,d} Illustration of time-of-flight images allowing the experimental observation of the twisted superfluid phase. The images correspond to the individual spin components of \textbf{a} and \textbf{b}, respectively. The alternating intensities of the first-order coherence peaks interchange between the species due to $\theta_\down\:-\theta_\up$. Here we chose $\theta$$\approx$$\pi/8$ and $A_\sigma\:1$, whereas at $\theta\:\pi/3$ the weaker peaks disappear entirely. \textbf{e} The relation between phase twist $\theta$ and spin-density wave for the infinitely degenerate ground state in the TSF phase. The markers correspond to 200 individual solutions of the self-consistent cluster mean-field algorithm with different random initial states.\label{fig_tof}}
\end{figure}

 The twisted superfluid phase for two components is characterized by anti-aligned twisted complex order parameters with $\theta_\up+\theta_\down\:0$ as illustrated in \Fig{fig_tof}a,b. The state is spontaneously symmetry-broken and forms degenerate pairs with $\theta_\up\:-\theta_\down\:\pm\theta$. This implies that the two components are in a strongly correlated state with respect to each other. The phase correlation is established by the cross tunneling process (equation \eqref{eq_H_2speciesUD}). Using the simple mean-field approach \eqref{eq_simple_mf}, the cross-tunneling energy is approximated by $2C \phi_{\uparrow i} \phi_{\downarrow j} \phi_{\downarrow i} \phi_{\uparrow j} \cos{(\theta_\uparrow - \theta_\downarrow)}$, which becomes minimal for $\theta_\uparrow \: - \theta_\downarrow \: \pi/2$. The actual values of $\theta_\uparrow$ and $- \theta_\downarrow$ are determined by all contributions in \eqref{eq_H_2species} and \eqref{eq_H_2speciesUD}. The cross tunneling energy for a correlated state  $C\expect{\hat a^\dagger_{\up i} \hat a^\dagger_{\down j} \hat a_{\down i} \hat a_{\up j}}$ can be large even in the vicinity of the Mott phases. In the pure mean-field approach of Ref.~\citenum{Choudhury2013}, it is pointed out that this beyond-mean-field effect could possibly yield a twisted superfluid phase. Our results show that the cross tunneling process is indeed mainly responsible for the emergence of the TSF phase at small amplitudes $P$. At the same time so-called spin-density waves may appear, which represent anti-aligned single-component density waves with $\Delta n_\up\:-\Delta n_\down$ such that the total density is homogeneous $\Delta n_\up$$+$$\Delta n_\down\:0$. 

The experimental observation of the density wave requires imaging with single-site resolution, whereas the complex twisting of the order parameter can be revealed in momentum space. Time-of-flight experiments with ultracold atoms allow the mapping on free momenta after a sudden release of the atomic trapping potential. For the individual spin components (we omit the spin index), the momentum density is proportional to the structure factor \cite{Gerbier2008} $S(\mathbf{k})=\sum_{ij}\mathrm{e}^{\I\mathbf{k}(\mathbf{r}_i-\mathbf{r}_j)} \expect{\hat a_i^\dagger \hat a_j }$, which gives six equivalent first-order momentum peaks $S_0$ at reciprocal lattice vectors for a superfluid ($\theta=0$). In the twisted superfluid, this symmetry is broken and adjacent peaks have either amplitude  $S_+$ or $S_-$. Figures~\ref{fig_tof}c,d depict calculated time-of-flight images yielding alternating patterns in momentum space. The two amplitudes are given by $S_\pm(\theta_\sigma)=2S_0[1+ A_\sigma \cos(\theta_\sigma\pm\pi/3)]$ for $\sigma$ either $\up$ or $\down$. For two components, the orientation of the stronger peaks are opposite for both components due to $S_\pm(\theta_\up)=S_\mp(\theta_\down)$. This behavior was also observed experimentally \cite{Soltan-Panahi2012}. An additional spin-density wave $\Delta n_\sigma$ modifies the strengths of the peaks via with $A_\sigma$, whereas $A_\sigma=1$ for a homogeneous density (see \Methods).

\subsection{Infinite degeneracy and quantum phase transitions}
In contrast to the two-fold degeneracy for a single species, we find that the TSF ground state for two components has an infinite degeneracy in respect to the continuous parameter $\theta$. More precisely, all states with $-\theta_\mathrm{max} \leq \theta\leq \theta_\mathrm{max}$ have the same energy, where $\theta$ can take a continuous value and $\theta_\mathrm{max}$ depends on the point in the phase diagram. As depicted in \Fig{fig_tof}e, the spin-density wave $\Delta n_\up\:-\Delta n_\down$ and the complex twisting $\theta$ are correlated. This corresponds to a redistribution of the energy between the density wave and the complex twisting of the superfluid order parameter. For $\theta\:\pm\theta_\mathrm{max}$ the spin-density wave vanishes, whereas the spin-density wave $|\Delta n_\sigma|$ is maximal for $\theta\:0$. These two extrema correspond to a "pure" twisted superfluid with $\Delta n_\sigma\:0$ and a real, anti-aligned supersolid in each component for $\theta\:0$, respectively. The latter implies that two of the solutions are real wave functions obeying time-reversal symmetry. The origin of the continuous symmetry lies within an equal redistribution of the energy between the different interaction processes and persists only if all amplitudes $V$, $P$ and $C$ are equal, which holds for equivalent atomic species with $SU(2)$ symmetric interactions. We observe this degeneracy for all studied cluster sizes. However, it might be lifted by including higher bands of the lattice leading to a renormalization of the amplitudes above\cite{Luhmann2012a}. At the touching points of TSF phases and CDW insulators, we observe a complex order parameter in combination with a total density wave $\Delta n_\up+\Delta n_\down\neq 0$, i.e.~a twisted supersolid (TSS). In analogy to the TSF phase, its ground state is continuously symmetry broken. We cannot exclude the possibility of a thin twisted supersolid layer at the boundary between supersolid and TSF phase.  

Similar, we find also an infinite degeneracy of the ground state for the MI and CDW insulator phases for symmetric intra- and interspecies interaction. For instance, the Mott insulator with filling $\expect{\hat n_{\up i} + \hat n_{\down i} } \:1$ on two neighboring sites $i$ and $j$ can be written in first order as $( c_1 \ket{\up}_i + c_2 \ket{\down}_i ) \otimes ( c_2 \ket{\up}_j - c_1 \ket{\down}_j )$ with an arbitrary constant $c_1$ and  $|c_1|^2+|c_2|^2=1$. Here, the intra- and interspecies interaction energy of nearest-neighbor sites $V$ and the cross tunneling energy $C$ are redistributed yielding the same total energy. Since the cross-tunneling order parameter $\langle \hat a_{\up i}^\dagger \hat a_{\down i} \rangle$ does not vanish, the two-species Mott insulator is sometimes referred to as a counterflow superfluid \citeCFSF.

As shown in \Fig{fig_pd2species}e, the transitions between SS and TSF phases are first order quantum phase transitions with discontinuities in $\partial\expect{\hat H}/\partial\mu$. At the transition point the total density wave $\Delta n_\up+\Delta n_\down$ and the superfluid order parameter are discontinuous. The structure of the ground state changes across the transition from a two-fold to infinite degeneracy. Also the transition between superfluid and Mott insulator are of first order \cite{Kuklov2004,Ozaki2012} characterized by a discontinuity of the superfluid order parameter as well as by the different degeneracies.   

The close-up of the phase diagram shown in \Fig{fig_pd2species}d reveals the appearance of dimerized density waves (DDW) for two components. These phases with quarter-integer filling arise between Mott and CDW insulators as discussed above for a single component. Due to the strong correlations between neighboring sites, this quantum phase is only found in calculations using clusters with at least two sites. Similar as the Mott insulator phase, the DDW has an infinite degeneracy. Attached to this phase, we also find a dimerized supersolid (DSS) in analogy to the single-component phase diagram.

\section{Discussion}
The appearance of the twisted superfluid phase at small values of the pair tunneling amplitude for a bosonic mixture is quite remarkable. For the honeycomb lattice a pair tunneling amplitude of $P/U=0.05$ and $0.01$ correspond to a lattice depth $V_0=1\ER$ to $3.5\ER$, respectively \cite{Luhmann2014}. The ratio $J/U$ depends on the scattering length $a_s$, where we find $J/U=0.14$ for $V_0=3.5\ER$ and $a_s=360a_0$. However, the ratios $P/J$ and $C/J$ can also be tuned experimentally using lattice shaking techniques \cite{Eckardt2005}. The presented results shed light on the formation of twisted superfluidity recently found experimentally \cite{Soltan-Panahi2012}. We are able to draw a consistent picture of the time-reversal symmetry breaking and the interplay with the insulating phases. The strikingly complex phase diagram is a manifestation of the multilayered interplay between different extended Hubbard processes on the one hand and nearest-neighbor correlations on the other hand. In the experimental realization \cite{Soltan-Panahi2012}, where the scattering length is about $a_s=100a_0$, the lattice potential depends on the atomic hyperfine states causing an energy offset between the sublattices. In future studies, the effect of the spin-dependent offset on the phase diagram can be evaluated.

Spontaneous breaking of the translational symmetry allows for dimerized density waves with quarter-integer filling. The nature of this fractional insulator is reflected in a strongly correlated superposition state on dimers of the lattice, which is linked to resonating valance bond coupling in solids. The same mechanism leads to dimerized supersolid phases. We have not found any indication for phase separation in the honeycomb lattice similar as Refs.~\citenum{Wessel2007,Gan2007a}. 
 
It would be interesting to study in detail the infinite degeneracy for the two-component TSF and TSS ground states. This degeneracy should be experimentally observable for a spin-independent lattice and $SU(N)$ symmetric interactions. For the hyperfine states of rubidiums atoms \cite{Soltan-Panahi2012}, the $SU(N)$ symmetry is broken causing slightly different intra- and interspecies scattering length. This also breaks the continuous ground-state degeneracy of the TSF phase and locks the value of $\theta$ and the density wave. However, the two-fold degeneracy connected with the time-reversal symmetry breaking persists.

For the sake of simplicity and to avoid an even larger parameter space, the Hamiltonians \eqref{eq_Hfull} and \eqref{eq_H_2species} are not complete with respect to nearest-neighbor interaction processes. They neglect a term of the form $ \hat a_i^\dagger(\hat n_i+\hat n_j) \hat a_i$ known as density-induced tunneling. For ultracold neutral atoms, its amplitude is comparably large. However, it has been shown that it may be absorbed in the single-particle tunneling $J$ in most situations \cite{Luhmann2012a,Maik2013,Dutta2015}. Thus, we mainly expect a shift of the phase boundaries due to this combined occupation-dependent tunneling. Moreover, the next-nearest-neighbor tunneling has a considerable amplitude in shallow lattices \cite{Luhmann2014}. Since its sign corresponds to antiferromagnetic coupling, it may even stabilize the twisted superfluid phase. Long-range interactions, such as dipolar interaction \cite{Dutta2015},  offer further perspectives for studying off-site interactions. 

 The twisted superfluid itself presents a highly exotic state of matter as it combines the off-diagonal long-range order of a superfluid state with a correlated short-range ordering. For two components, correlations between the species are essential and manifest themselves in spin-density waves and an opposite complex twisting of the superfluid order parameter. As previously mentioned the connection to time-reversal symmetry breaking in the context of superconductors \cite{Kaminski2002, Luke1998, Sauls1994, Strand2010, Schemm2014} is of high interest. Quantum gas experiments may contribute by offering highly tunable systems and a different experimental platform for studying correlated quantum phases in lattices. 

\hypertarget{MethodSection}{}\section{Methods} 
\methodsection{Cluster mean-field method and order parameters}
For the computation of the phase diagram, we apply the cluster Gutzwiller method. It is based on factorizing the many-body wave function $\ket{\psi}\ket{\Psi_\mathrm{c}}$ in two parts describing particles inside $\ket{\Psi_\mathrm{c}}$ and outside the cluster $\ket{\psi}$. Accordingly, the Hamiltonian decomposes in $\hat H_{NM} = \bra{M} \hat H_\text{cluster} + (\bra{\psi}\hat H_\text{boundary}\ket{\psi}) \ket{N}$ in the basis of many-particle cluster states $\ket{N}$. The knowledge of the (factorized) solution for $\ket{\psi}$ allows obtaining the correlated cluster wave function $\ket{\Psi_\mathrm{c}}=\sum_N c_N \ket{N}$ by means of diagonalization. Fortunately, tight-binding Hamiltonians reduce the information required from the wave function $\ket{\psi}$ outside the cluster, since $\hat H_\text{boundary}$ acts only on sites at the boundary.  For the  single-component  Hamiltonian \eqref{eq_Hfull}, $\bra{\psi}\hat H_\text{boundary}\ket{\psi}$ reduces to $\expect{\hat a_i }$, $\expect{\hat n_i}$ and $\expect{\hat a_i \hat a_i}$, which we obtain self-consistently from inside the cluster. Note that $\hat H_\text{boundary}$ couples different Fock spaces in the cluster. In the case of the two-component Hamiltonian \eqref{eq_H_2species} eight mean-field parameter are required, i.e. 
$\expect{\hat a_{\up i} }$, $\expect{\hat n_{\up i}}$, $\expect{\hat a_{\up i} \hat a_{\up i}}$, 
$\expect{\hat a_{\down i} }$, $\expect{\hat n_{\down i}}$, $\expect{\hat a_{\down i} \hat a_{\down i}}$, 
$\expect{\hat a_{\up i} \hat a_{\down i}}$ and $\expect{\hat a_{\up i}^\dagger \hat a_{\down i}}$. 
We apply self-consistent iterations $I$, until $\sum_N |c_{N,I}-c_{N,{I-2}}|$ falls below a threshold value (typically smaller than $10^{-8}$). On subsequent iterations, we allow the cluster wave function and mean fields to be different, which enables us to find structures spanning $2k$ sites with a $k$-site cluster. Due to the high dimension of the mean-field parameter space up to several thousand iterations are needed in certain regions of the phase diagram. We have carefully checked convergence with respect to the size of the cluster basis. Further technical details can be found in Ref.~\citenum{Luhmann2013}.

\methodsection{Phase diagrams}
The phase diagrams presented in Figs.~\ref{fig_pd} and \ref{fig_pd_cut} are obtained by means of the cluster mean-field approach using a two-site cluster. The phase boundaries are slightly affected when increasing the cluster size (see comparison in \Fig{fig_pd_cut}a), since longer-ranged many-particle correlations can be incorporated \cite{Luhmann2013}. The effective chemical potential $\mu_\eff = (\mu + U/2)/(1+6P/U) -U/2$ accounts for the increased total interaction energy for finite $V$. In these units, the position of the Mott lobes is approximately independent of $P\:V$. The phase diagram for two components in \Fig{fig_pd2species}a,b is computed using a single-site cluster and the two-component tensor basis, i.e. including all interspecies on-site correlations. The close-ups in \Fig{fig_pd2species}c,d are obtained using two-sites clusters. For two components, the chemical potential $\mu_\eff = (\mu + U/2)/(1+3P/U) -U/2$ accounts for the off-site interaction energy.

\methodsection{Simple variational mean-field expressions}
For a single component, a variational model can be constructed serving as a motivation for the complex phase in the twisted superfluid phase. In contrast to the cluster mean-field approach above, we neglect all correlations arising from the one- and two-particle operators by setting $\expect{\hat a_i^\dagger \hat a_j} \: \expect{\hat a_i}^*\expect{\hat a_j}$ and $\expect{\hat a_i^\dagger \hat a_j^\dagger \hat a_k \hat a_l} \: \expect{\hat a_i}^*\expect{\hat a_j}^*\expect{\hat a_k}\expect{\hat a_l}$. Allowing for two different sublattices A and B, we replace the operators with complex numbers  $\expect{\hat a_A} \: \phi_\mathrm{A}$ and $\expect{\hat  a_B} \: \phi_\mathrm{B} \e^{\I \theta}$ with a relative phase $\theta$. In this model, the mean-field energy of Hamiltonian \eqref{eq_Hfull} per unit cell reads $\expect{H}=U(\phiA^4$$+$$\phiB^4)/2 -\mu (\phiA^2$$+$$\phiB^2) - 6J \phiA\phiB \cos(\theta)+ 6 V \phiA^2\phiB^2	+ 3 P   \phiA^2\phiB^2 \cos(2\theta)$. This model predicts the transition between superfluid and twisted superfluid qualitatively, but is unable to predict correlated insulator phases and is invalid for small $J/U$. The same model can also be applied to the two-species case  (equation \eqref{eq_H_2species}) with the same restrictions mentioned above.  In particular, this approach cannot capture the twisted superfluid phase at small amplitudes of the off-site interactions, which is driven by correlations. 

\methodsection{Structure factor}
The twisted superfluid state gives rise to a distinct time-of-flight signature showing first-order momentum peaks with an alternating intensity $S_\pm$ (\Fig{fig_tof}). A density wave $\Delta n_\sigma$ within the spin component $\sigma$  modifies the strengths of the peaks $S_\pm$ via $A_\sigma=(1-\Delta n_\sigma^2/N^2)^{1/2}$ with $N=\sum_\sigma \expect{\hat n_{\sigma,i}+ \hat n_{\sigma,j}}$. The central peak for zero momentum is proportional to $1+A_\sigma \cos(\theta_i)$.  

\section{References}\vspace{-1.6cm}

\subsection{Acknowledgements}
We thank J.~Struck as well as L.~Cao, S.~Kr{\"o}nke, J.~Stockhofe and P.~Schmelcher for stimulating discussions. 
We acknowledge funding by the Deutsche Forschungsgemeinschaft (grants SFB~925 and the The Hamburg Centre for Ultrafast Imaging (CUI)). 

\end{document}